\documentclass[english,aps,manuscript]{revtex4}
\usepackage[T1]{fontenc}
\usepackage[latin9]{inputenc}
\usepackage{float}
\usepackage{graphicx}

\makeatletter
\@ifundefined{textcolor}{}
{%
 \definecolor{BLACK}{gray}{0}
 \definecolor{WHITE}{gray}{1}
 \definecolor{RED}{rgb}{1,0,0}
 \definecolor{GREEN}{rgb}{0,1,0}
 \definecolor{BLUE}{rgb}{0,0,1}
 \definecolor{CYAN}{cmyk}{1,0,0,0}
 \definecolor{MAGENTA}{cmyk}{0,1,0,0}
 \definecolor{YELLOW}{cmyk}{0,0,1,0}
}

\makeatother

\usepackage{babel}
\begin{document}

\title{Heavy and Light Monopoles in Magnetic Reversion in Artificial Spin
Ice}

\author{Alejandro León}

\address{Facultad de Ingeniería, Universidad Diego Portales, Santiago Chile}
\begin{abstract}
This work makes a theoretical study of the dynamics of emergent elemental
excitations in artificial spin ice systems with hexagonal geometry
during the magnetic reversion of the system. The magnetic and physical
parameters of the nanoislands that form the array are considered as
variables in the study. The parameters considered are: the energy
barrier for the inversion of each nanoisland, the magnetic moment
of the nanomagnets and the possible disorder in the sample. Our results
show that the reversion dynamic presents two distinct mechanisms of
magnetic reversion, with different elemental excitations for each
mechanism. The first mechanism presents a reversion with the appearance
of magnetic monopoles that do not move in the samples (heavy monopoles)
and the absence of Dirac chains. In the other mechanism elemental
magnetic excitations (light monopoles) appear that move great distances
in the sample, giving rise to extensive Dirac chains during the magnetic
reversion.
\end{abstract}
\maketitle

\section*{Introduction}

Artificial spin ice systems have been intensely studied theoretically
and experimentally in recent years {[}1-9{]} because they present
magnetic charge defects with similar characteristics to the magnetic
monopoles postulated by Dirac {[}10{]}. The main advantage of these
systems lies in the fact that studies can be made of the dynamic of
these elemental excitations at ambient temperature because the nanoislands
have energy reversion barriers above the thermal fluctuations at this
temperature. The work of Ladak et al. {[}7{]} showed the direct observation
of these elemental excitations in a cobalt nanoisland array with hexagonal
geometry. Another recent study {[}8{]} showed the experimental evidence
of magnetic reversion through Dirac chains in an artificial spin ice
system with hexagonal geometry. The work of Mengotti et al. {[}8{]}
includes the statistics that demonstrate the decrease in the dimensionality
of the system based on the propagation of Dirac chains, with mobile
elemental excitations (monopoles) at their ends. Another recent experimental
work {[}9{]} verified that magnetic reversion is produced independently
in a square network for the two sub-networks that form the array.
As well, the same study verified that the reversion is produced through
unidimensional cascades of the parallel nanomagnets of the system.
Some studies have been published on the effect of the size of the
system in artificial spin ice arrays on energy minimization protocols
{[}11{]} and the effect of the edges on vertice dynamics {[}12{]}.
A recent study {[}13{]} was published on the effects of the size of
finite systems, the aspect ratio of the system and the concentration
of impurities on the dynamic of emergent monopoles in a hexagonal
network during magnetic reversion. This work establishes that the
density of mobile monopoles depends on the size, concentration of
impurities and shape of the bidimensional system.

The current techniques of synthesizing artificial spin ice systems
{[}1, 7-9{]} allows for controlling the size and shape of the system
and the magnetic properties of the individual nanoislands. However,
there is still no fine control of the relationship between the magnetic
moments of each nanoisland and the external magnetic field necessary
to provoke magnetic reversion in isolated nanoislands, giving rise
to the possibility of having a broad spectrum of combinations of these
two variables. In this work we studied the dynamic of magnetic reversion
in a hexagonal array of nanomagnets in function of the magnetic properties
of the individual components of the array. Our results show that there
are two different reversion mechanisms with two distinct types of
emergent magnetic monopoles, depending on the energy barrier for the
reversion of the nanomagnets and on their magnetic moment. In one
case monopoles emerge and remain in repose during the reversion. We
term this type of excitation \textquotedblleft{}heavy monopoles\textquotedblright{}.
In the other case, elemental excitations emerge that move great distances
through the sample generating extensive Dirac chains. We term this
type of excitation \textquotedblleft{}light monopoles\textquotedblright{}.

\section*{Emergent Magnetic Monopoles}

The system that we studied is a bidimensional array of magnetic nanoislands
on a hexagonal network. Three nanomagnets converge in each vertice
of the array (except on the edges). We assign a magnetic load value
to each vertice of the array in function of the magnetic poles present
in the vertice. The north pole of each nanoisland is associated with
a $q=+1$ charge and the south pole is associated with a $q=-1$ charge.
This implies that each vertice of the array has the following net
charge values: $Q_{1}=+1$, $Q_{2}=-1$, $Q_{3}=+3$ and $Q_{3}=-3$.
If the system is submitted to a magnetic field along the $x$-axis,
as shown in the upper part of Figure 1, the vertices in the array
(except at the edges) acquire a charge of $+1$ or $-1$. If the samples
are initially magnetized, to the left, for example, and we apply a
magnetic field in the inverse sense, elemental excitations appear
(magnetic monopoles) in the vertice. The monopoles are defined as
in reference {[}8{]}. When the samples are totally magnetized to the
left, a positive monopole emerges if the net charge of the vertice
\textquotedblleft{}A\textquotedblright{} goes from $Q=-1\rightarrow Q'=+1\Rightarrow\triangle Q=+2$.
Equally, a negative monopole results if vertice \textquotedblleft{}A\textquotedblright{}
goes from $Q=-1\rightarrow Q'=-3\Rightarrow\triangle Q=-2$. In the
case of the type B vertices, a negative monopole emerges if $Q=+1\rightarrow Q'=-1\Rightarrow\triangle Q=-2$
and a positive monopole emerges when $Q=+1\rightarrow Q'=+3\Rightarrow\triangle Q=+2$.
The monopoles emerge in pairs and then move away from each other but
connected through Dirac chains, as a product of the applied magnetic
field, as can be observed in the lower part of figure 1. A possible
combination could occur if  $\triangle Q=\pm4$, but these events
have not been observed in experiments or simulations. Of the emergent
monopoles only those that maintain at least one of the nanomagnets
that converge in the vertice are mobile, with their component over
the $x$-axis directed to the left. We termed the total quantity of
mobile monopoles for each site of the network the \textquotedblleft{}density
of mobile monopoles\textquotedblright{} with the designation  $\sigma_{M}$.
When the conditions  $\triangle Q=\pm2$, are present in a vertice,
but with all the nanomagnets, that converge in the vertice, with the
component of the magnetic moment directed to the right, monopoles
emerge that are termed \textquotedblleft{}trapped monopoles\textquotedblright{}
{[}8{]} and in this work are not considered as monopoles properly
but rather as \textquotedblleft{}charge defects\textquotedblright{}.
We termed the quantity of these defects per site of the hexagonal
network as \textquotedblleft{}defect density\textquotedblright{} and
denote it by  $\rho$. 

\begin{figure}[h]
\includegraphics[scale=0.25]{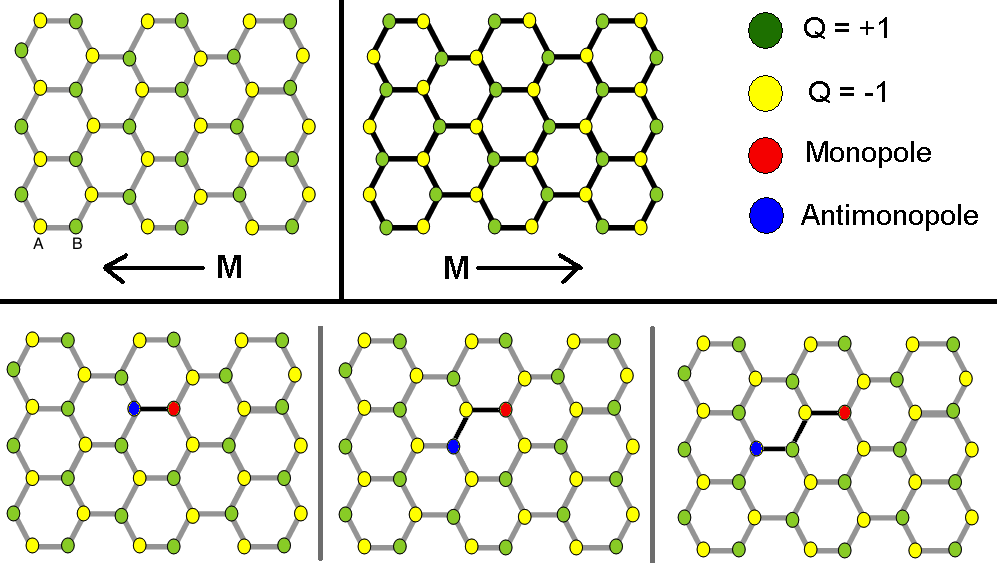}

Figure 1. The upper left part of the figure shows the array of nanoislands
magnetized to the left and the upper middle part of the figure shows
the array of nanoislands magnetized to the right. All the nanoislands
that have a negative value for the $x$ component of the magnetic
component are represented by gray lines. In the case that the $x$
component of the magnetic moment of the nanoisland is positive, it
is represented by a black line. The upper right part of the figure
shows the color code for the vertices and the emergent monopoles.
\end{figure}

\section*{The studied system and the model used}

The system studied, is a sample of $\left(23\;\mu m\times23\;\mu m\right)$
and 1,144 nanomagnets, with impurities. The lattice constant (the
distance between two adjacent vertices) has a value of $a=577\: nm$.
The random magnetic moment of individual islands is given by $m=m_{0}\beta$,
where $\beta$ is a dimensionless Gaussian random variable with $\left\langle \beta\right\rangle =1$
and $s\equiv\left(\left\langle \left(\beta-\left\langle \beta\right\rangle ^{2}\right)\right\rangle \right)^{1/2}$.
We studied the dynamic of emergent monopoles during magnetic reversion
in function of the $m_{0}$ moment of the individual nanoislands;
of $H_{b}$, the necessary field to revert the moment of the individual
nanoislands and of $s$, the concentration of impurities in the sample.
Using a model based on a frustrated cellular automata (FCA), we consider
the dipolar interaction between the magnetic moments in the Hamiltonian
of the system and the dynamic evolution of the system. The mathematical
basis of the model and its applications are explained in detail in
reference {[}14{]}. Likewise, the results of reference {[}13{]} validate
our model using experimental data from the work of Mengotti et al.
{[}8{]}. In the FCA, the automaton is updated in several steps (three
in this case), minimizing the total energy total of the system, evaluated
with dipolar interaction.

\section*{Heavy and light monopoles}

The methodology used in our work consists of fixing the magnetic field
$H_{b}$, necessary to revert an isolated nanoisland and simulate
magnetic reversions of the array for different values of  $m_{0}$.
Just one simulation provided us with information of the hysteresis
curve, the total density of impurities $\rho$ and the density of
mobile monopoles in function of the applied magnetic field. During
the magnetic reversion of the system, the quantity of mobile monopoles
increases, reaches a maximum, for a field value close to the coercivity
field and then decreases to zero. As relevant data of the simulation,
we use the maximum value reached of the density of the $\sigma_{M}$
mobile monopoles. We term this maximum value $\sigma_{max}$. Because
we simulated the presence of impurities in the system, with a random
procedure we repeated the simulation for each experiment 100 times
and determined  $\left\langle \sigma_{max}\right\rangle $ and the
standard error of the measured samples. Figure 2 shows the results
for four values of the reversion field of the nanoislands. Each graph
shows the maximum value of the average density of the mobile monopoles
in function of the magnetic moment $m_{0}$ of the nanoislands. In
each case, three concentrations of impurities are considered. We can
appreciate two distinct behaviors in all the graphs. A first region,
dominated by the presence of heavy monopoles, where the magnetic reversion
occurs with the appearance of elemental excitations, but these do
not move in the sample and remain static. In this region, the samples
with few impurities present a higher number of monopoles, reaching
100\% in 3 situations shown in Figure 2. This is explained by the
fact that the disorder in the system allows that diagonal nanoislands
join two horizontal nanoislands that have heavy monopoles in their
extremes and thus the number of these monopoles decreases without
the movement of these excitations. This situation is shown in Figure
4.1. When the system has a low percentage of impurities, the probabilities
of these unions is very low and consequently all the sites of the
network end up with a heavy monopole. Similarly, we can note in Figure
2 that there is a threshold value for the magnetic moment $m_{0}$.
This threshold value separates the region of heavy monopoles from
the region of light monopoles. Monopoles are generated in the light
excitation region that can move large distances in the system, extending
the Dirac chains that join the monopole pair created, which is presented
in Figure 4.3. The impurities in this region play a distinct role
from the previous case, given that it permits the appearance of more
monopole pairs and consequently increases $\sigma_{max}$. density.
The upper right of Figure 2 shows the results with the experimental
parameters used in validating the model {[}13{]}, with $m_{0}=1.1\times10^{-15}\: A\cdot m^{2}$
the experimental value for nanoisland moment. 

\begin{figure}[h]
\includegraphics[scale=0.5]{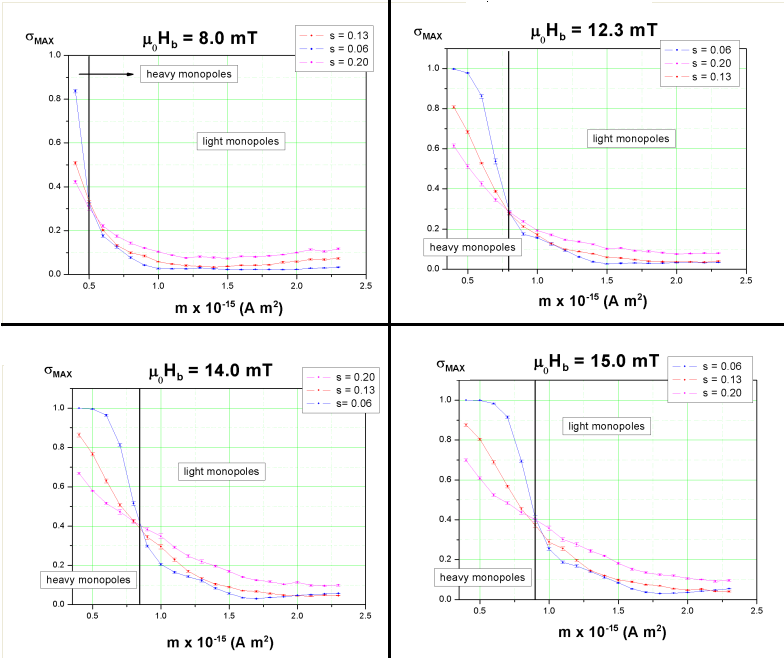}

Figure 2. Maximum density of mobile monopoles in function of the magnetic
moment of the nanoislands that form the array. The graphs represent
four field values to revert the magnetic moment of the nanoislands.

\end{figure}

Figures 3 and 4 show simulations with these experimental parameters,
except for $m_{0}$. Figure 4.1 shows the situation in the case of
heavy monopoles $m_{0}=0.4\times10^{-15}\: A\cdot m^{2}$, Figure
4.2 shows the reversion in the threshold value of the magnetic moment
of the nanoislands $m_{0}=0.8\times10^{-15}\: A\cdot m^{2}$ and Figure
4.3 shows reversion in the light monopoles regime $m_{0}=1.5\times10^{-15}\: A\cdot m^{2}$.
We can note in Figure 4.2 that the magnetic reversion in the threshold
value is produced with the coexistence of heavy and light monopoles.

Figure 3 shows the hysteresis curve, the maximum density of monopoles
and the density of defects for the simulations, which is shown in
Figure 4. We can note that coercivity is reached in the heavy monopole
regime before it is in the light monopole regime. In the light monopole
regime magnetization goes abruptly from a value of -1 al to 0.5. It
is interesting to note that this characteristic could be desirable
to use this type of system to propagate binary information {[}15{]}.
A species of plateau is then produced for the three curves.

\begin{figure}
\includegraphics[scale=0.35]{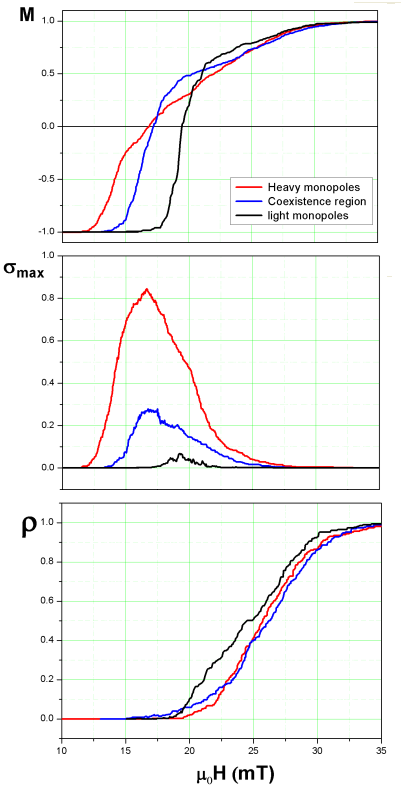}

Figure 3. Magnetization $M$, maximum density of $\sigma_{max}$ monopoles
and the density of defects $\rho$ for the three simulations shown
in Figure 4. The red curves correspond to the simulation shown in
Figure 4.1, the blue curves correspond to Figure 4.2 and the black
to Figure 4.3.

\end{figure}

\begin{figure}[H]
\includegraphics[scale=0.45]{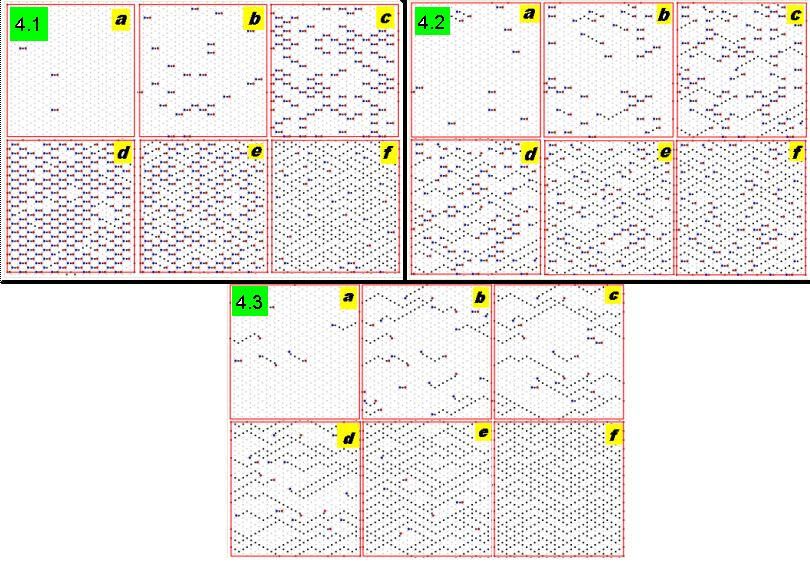}

Figure 4. Simulation of the magnetic reversion for the parameters
$\mu_{0}H_{b}=12.3\: mT$, $s=0,13$, of the experimental work {[}8{]},
except for the magnetic moment. Figure 4.1 shows a simulation in the
heavy monopole regime, $m_{0}=0,4\times10^{-15}\; A\cdot m^{2}$,
Figure 4.2 shows the region of coexistence of heavy and light monopoles,
$m_{0}=0,8\times10^{-15}\; A\cdot m^{2}$ and Figure 4.3 shows light
monopoles regime, $m_{0}=1,5\times10^{-15}\; A\cdot m^{2}$. Each
simulation shows the central region of the nanoisland array during
the magnetic reversion for six external magnetic field values. The
gray squares represent nanoislands with the $x$ component of the
magnetic moment directed to the left and the black squares to the
right. The red circles represent positive monopoles and the blue circles
negative monopoles. In the three situations, figures a, b and c correspond
to values of the external magnetic field, lower than the coercivity
field  $H<H_{C}$. Figure d represents the case for $H=H_{C}$ and
figures e and f represent situations in which $H>H_{C}$.

\end{figure}

In our results, we can observe that the length of the Dirac chains
present in the reversion is inversely proportional to the density
of the mobile monopoles. This implies that in the region of light
monopoles, impurities do not favor the formation of extended Dirac
chains. In this region, the samples with few impurities show a rich
dynamic in very extensive Dirac chains and two distinct reversion
phases. In the first phase, monopoles appear, generally at the ends
and moving to the opposite extremes. These first monopoles disappear,
producing a first plateau in the hysteresis curve. Then for a magnetic
value close to coercitivity, new monopoles and the corresponding Dirac
chains are produced. The second chains are shorter than the first.
A second plateau is produced and then the samples reach total magnetization.
This behavior can be observed in Figure 5, which shows the magnetization
for a simulation with the same parameters as shown in Figure 4, but
in the limit case where there are no impurities $\left(s=0\right)$.
The same figure shows the heavy monopole regime without impurities.
We can note that when the system reaches the coercitivity value, it
remains in an extended plateau. During this plateau, the density of
the mobile monopoles is at a maximum. There is then a second abrupt
rise and a second plateau. In the final part of the magnetic reversion
there is a third abrupt rise, reaching the maximum value of magnetization.
It is only in the last rise that the density of defects goes from
a low to a maximum value.

\begin{figure}[H]
\includegraphics[scale=0.25]{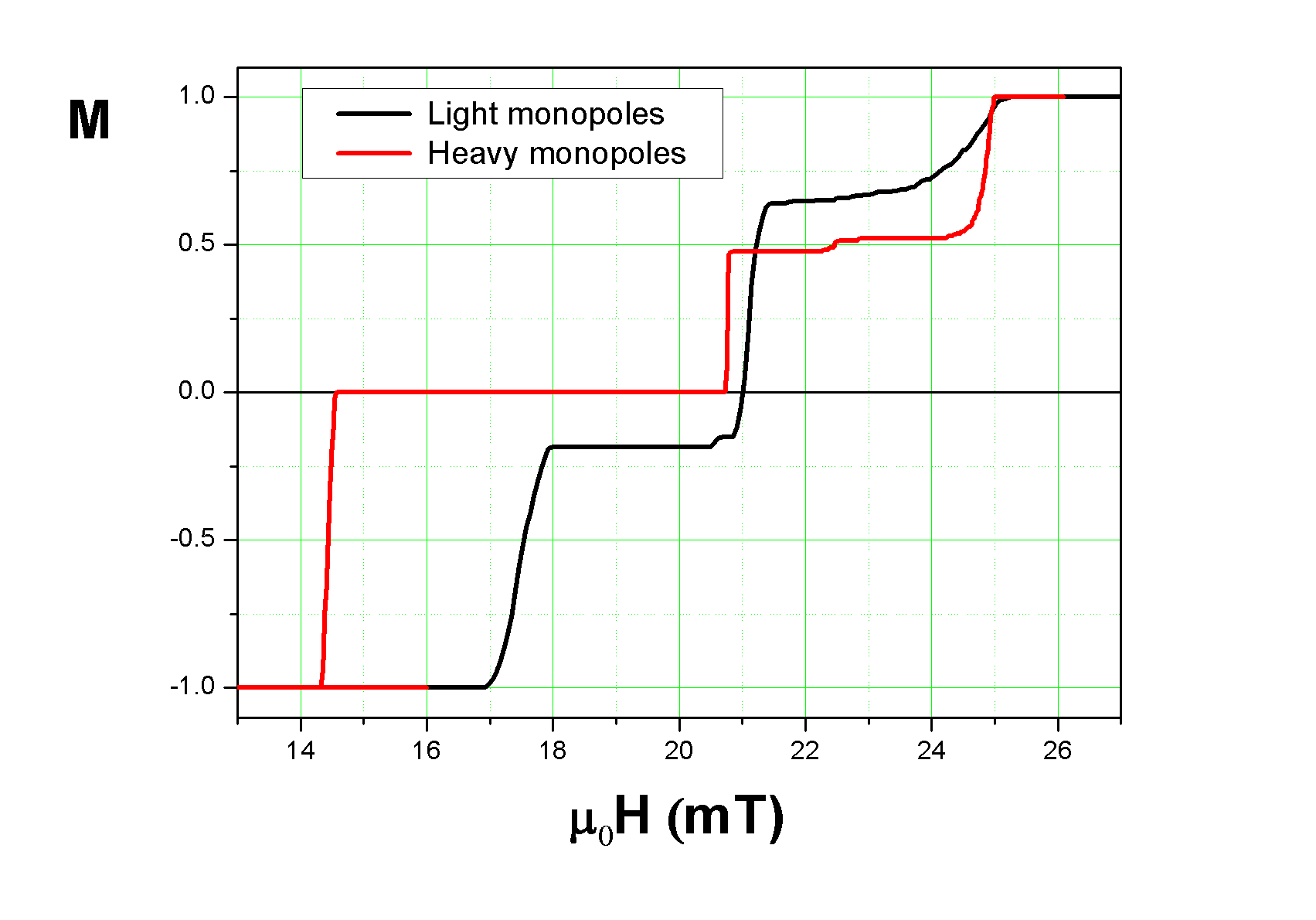}

Figure 5. Magnetization for a simulation with the same parameters
as shown in Figure 4, but in the limit case where there are no impurities
$\left(s=0\right)$.

\end{figure}

\section*{Conclusion}

Magnetic reversion in artificial spin ice systems with hexagonal geometry
presents two different dynamic mechanisms of elemental excitations,
depending on the magnetic properties of the individual nanoislands
that form the array. In one case we can appreciate that reversion
occurs with the appearance of a large number of monopoles that however
remain static in the sample. In the other case, we can note the low
number of monopoles that move great distances in the samples with
very extensive Dirac chains. Impurities in the sample condition the
number of these heavy and light monopoles. In the heavy monopole regime,
the impurities decrease the number of monopoles and in the light monopole
regime the opposite occurs, increasing the number of monopoles.

\section*{Acknowledgments}

The author acknowledge the financial support of FONDECYT program grant
11100045.

\end{document}